# Tests of the continuum limit for the $SO(4)$ Principal Chiral Model and the prediction for $\Lambda_{\overline{MS}}$

**M. Hasenbusch and R.R. Horgan**

D.A.M.T.P.
Silver Street, Cambridge, England CB3 9EW

November 2, 1995

## Abstract

We investigate the continuum limit in $SO(N)$ Principal Chiral Models concentrating in detail on the $SO(4)$ model and its covering group $SU(2) \otimes SU(2)$. We compute the mass gap in terms of $\Lambda_{\overline{MS}}$ and compare with the prediction of Hollowood [1] of $m/\Lambda_{\overline{MS}} = 3.8716$. We use the finite-size scaling method of Lüscher et al. [2] to deduce $m/\Lambda_{\overline{MS}}$ and find that for the $SO(4)$ model the computed result of $m/\Lambda_{\overline{MS}} \sim 14$ is in strong disagreement with theory but that a similar analysis of the $SU(2) \otimes SU(2)$ yields excellent agreement with theory. We conjecture that for $SO(4)$ violations of the finite-size scaling assumption are severe for the values of the correlation length, $\xi$, investigated and that our attempts to extrapolate the results to zero lattice spacing, although plausible, are erroneous. Conversely, the finite-size scaling violations in the $SU(2) \otimes SU(2)$ simulation are consistent with perturbation theory and the computed $\beta$-function agrees well with the 3-loop approximation [2] for couplings evaluated at scales $L/a \leq \xi$, where $\xi$ is measured in units of the lattice spacing, $a$. We conjecture that lattice vortex artifacts in the $SO(4)$ model are responsible for delaying the onset of the continuum limit until much larger correlation lengths are achieved notwithstanding the apparent onset of scaling. Results for the mass spectrum for $SO(N)$, $N = 8, 10$ are given whose comparison with theory gives plausible support to our ideas.

manifold $M = G/H$, where $G$ and $H$ are Lie groups and $H \subset G$, has provided a deep understanding of field-theoretic methods applied to both high energy physics and critical phenomena. In particular, the continuum field theories derived from these models are renormalizable and asymptotically free and so act as good models to test ideas applicable to four-dimensional gauge theories. For many models the exact S-matrix for the continuum field theory has been conjectured [3, 4, 5] and this allowed a number of workers [5, 6, 7] to give exact predictions for the mass spectrum and to use the Thermodynamic Bethe Ansatz together with the property of asymptotic freedom to predict the ratio $m/\Lambda_{\overline{MS}}$ for these theories, where $m$ is mass-gap and $\Lambda_{\overline{MS}}$ sets the mass scale in perturbation theory. It is important to note that solely the local properties of $M$, i.e., the algebra which generates infinitesimal displacements on $M$, and not the global properties of $M$ are used to obtain these results. To be able to control the approach to the continuum limit for a lattice field theory is central to current calculations in lattice QCD where criteria based on scaling are used to determine whether dimensionless ratios for computed observables have attained their continuum values and so can be compared with experiment. The existence of theoretical predictions for 2D chiral field theories which link the short and long range structure of the theories gives a unique opportunity to study the approach to the continuum in a lattice model and also to test the assumptions underlying the theoretical analysis.

The correlation length is most conveniently measured in units of the lattice spacing, $a$, and is denoted by the dimensionless variable, $\xi$, and the mass-gap is defined by $m = 1/a\xi$. For $\xi$ sufficiently large the mass spectrum inferred from the correlation functions of suitable chosen interpolating operators and the mass-gap measured in units of $\Lambda_{\overline{MS}}$ should agree with theory. Such agreement has been found for the $SU(N)$ matrix models [8, 9] for values of $N = 3 - 15$ already for moderately small correlation lengths: $\xi > 5$. However, the manifold of $SU(N)$ is simply connected but this is not so for many other models for which it is important to see if any obstructions/lattice artifacts delay the onset of the continuum limit. We may then learn how to be more confident in controlling the approach to the continuum. Similarly good agreement with theory has also been found by Wolff [10] for the $O(4)$ and $O(8)$ spin models but not for the $O(3)$ spin model [11, 12]. A comparison of $O(3)$ and $RP^2$ models has recently been the subject of much attention [13, 14, 15, 16, 17] since they are expected to correspond to the same continuum theory although they differ in the global properties of the defining manifolds. It has, however, proved impossible to demonstrate the equivalence of these two models in the continuum by computer simulation alone. The study reported in this paper highlights the difficulty of reliably controlling the approach to the continuum in a number of common lattice models.

We study the $SO(N)$ matrix models on the lattice with the action given by

$$S(\mathbf{U}) = \beta_L \sum_{x,t} Tr\left(\mathbf{U}(x,t)\mathbf{U}^T(x,t+1) + \mathbf{U}(x,t)\mathbf{U}^T(x+1,t)\right), \qquad (1.1)$$

where $\mathbf{U}(x,t)$ is an $N \times N$ orthogonal matrix. Our analysis mainly uses the finite size scaling method described by Lüscher et al. in reference [2] which is based on the coupling introduced by Nightingale [18]. For the lattice regularized theory we define the coupling constant $\tilde{u}$ by

$$\tilde{u}(L/a, \beta_L) = m(L/a, \beta_L)L, \qquad (1.2)$$

length is determined by the decay of the correlation function on the $L \times T$ strip in the time direction. The physical properties can be separated from the cutoff dependence and the continuum coupling, $u(Lm_\infty)$, can be defined by writing

$$\tilde{u}(L/a, \beta_L) = \bar{u}(Lm_\infty, a/L) = u(Lm_\infty) + O((a/L)^\omega) . \qquad (1.3)$$

The physical scale is set by $m_\infty = 1/a\xi_\infty$, the mass-gap in the infinite-volume limit. Apart from a trivial reparametrization, this decomposition relies on the physical hypothesis of scaling with the exponent $\omega$ parametrizing the corrections to scaling. For sufficiently large $\beta_L$ and sufficiently small $L/a$ perturbation theory can be used to compute $u$, and since it is proportional to $1/\beta_L$ at tree-level it can be viewed as a definition for a running coupling of the continuum field theory. In particular $u$ will evolve according to the universal part of the $\beta$-function.

$u(Lm_\infty)$ can be calculated from a simulation of the lattice theory. For $Lm_\infty \geq 1$ this is straightforward: for a given value of $\beta_L$ compute $am_\infty$ and then simulate on a lattice of width $L$ given in units of $a$ by $L/a = Lm_\infty/am_\infty$, and so compute $\tilde{u}(L/a, \beta_L)$ from equation (1.2). Repeat for a sequence of $\beta_L$ values, the largest of which is determined by computer resources. Check whether the results are compatible with the scaling hypothesis (equation (1.3)) and, if so, take $\bar{u}(Lm_\infty, a/L)$ with the largest $L/a$ as the best approximation to $u(Lm_\infty)$ or, better still, use equation (1.3) to fit the $a$-dependence and extrapolate to $a = 0$.

However, this direct approach cannot be used for small values of $Lm_\infty$ since $L/a \geq 1$ and $am_\infty$ is bounded from below by computer resources and so limits how small $Lm_\infty$ can be. This problem is resolved using the renormalization group by computing the lattice step-scaling function $\Sigma$:

$$\Sigma(s, \tilde{u}(L/a, \beta_L)) = \tilde{u}(sL/a, \beta_L) . \qquad (1.4)$$

Since no reference is made to $m_\infty$ we can choose, for any given value of $\tilde{u}$, $L/a$ as large as the computer resources will allow by tuning $\beta_L$ appropriately. Again we expect the scaling hypothesis to hold:

$$\Sigma(s, \tilde{u}(L/a, \beta_L)) = \sigma(s, u) + O((a/L)^\omega) . \qquad (1.5)$$

$\sigma(s, u)$ is the continuum step-scaling function determined implicitly from the $\beta$-function by

$$\log(s) = -\int_u^{\sigma(s,u)} \frac{du'}{\beta(u')} . \qquad (1.6)$$

The $\beta$-function has the perturbative expansion

$$\beta(u) = -u^2 \sum_{l=0}^\infty b_l u^l , \qquad (1.7)$$

Having computed $u(2)$, say, by the direct method a sequence of values $u_n \equiv u(2s^n)$ can be computed using

$$u_n = \sigma(s, u_{n-1}) . \qquad (1.8)$$

A reasonable choice for $s$ is $s = 1/2$. For couplings, $u_n$, evaluated at sufficiently large $n$ $\sigma(s, u)$ is well approximated by perturbation theory and $\Lambda_\xi$ is given to 2-loop order by

$$L\Lambda = (b_0 u(Lm_\infty))^{-b_1/b_0^2} \exp\left(-\frac{1}{b_0 u(Lm_\infty)}\right) , \qquad (1.9)$$

to overall rescaling) and for some cases $b_2$ is known [2] yielding a 3-loop estimate for $\Lambda_\xi$. A 1-loop calculation then relates $\Lambda_\xi$ to $\Lambda_{\overline{MS}}$.

Alternatively, the approach used in references [12, 10, 8, 9] can be employed where a bare lattice coupling is defined to be a running coupling evaluated at the scale given by the lattice spacing, $a$. Two candidates are

$$u_L(am_\infty) = \frac{1}{\beta_L} : \qquad \text{the 'lattice scheme'},$$
(1.10)
$$u_E(am_\infty) = \frac{4\langle E \rangle}{C} : \qquad \text{the 'energy scheme'}, \qquad (1.11)$$

where $C = n_f$, $n_f/N$ for spin and matrix models respectively, and $n_f$ is the number of degrees of freedom in the field. The coupling $u_E$ was first defined by Parisi in reference [19]. For $\xi_\infty \gg 1$ both of these couplings should scale with $a$ according to the perturbative $\beta$-function thus allowing $a\Lambda_L$ and $a\Lambda_E$ to be computed from $u_L$ and $u_E$ respectively according to equation (1.9) with $a$ replacing $L$. Since $\xi_\infty$ is known this allows $m/\Lambda_{\overline{MS}}$ to be estimated once the 1-loop calculations relating $\Lambda_{\overline{MS}}$ to $\Lambda_L$ and $\Lambda_E$ have been done.

The main motivation for our study was the recent work by Hollowood [1, 20] in which he has given theoretical predictions for $m/\Lambda_{\overline{MS}}$ for all $N > 3$, so allowing us to compare our lattice calculations with theory. The comparison between the studies of the lattice and continuum models sheds light on the approach to the continuum limit in the lattice formulation and can provide verification of the assumptions underlying the theoretical analysis of the continuum theory. We calculate $\bar{u}(Lm_\infty, a/L)$ defined in equation (1.3) by simulation for various values of $am_\infty$ and discuss the extrapolation of the results to $a = 0$ and we repeat the analysis for the covering group of $SO(4)$, namely $SU(2) \otimes SU(2)$. We also compare the results with an analysis based on the lattice– and energy–schemes for the couplings defined in equation (1.11). For the $SO(4)$ model we find no evidence that finite-size scaling holds for the accessible range of couplings, but that the results can deceptively suggest that it does hold since the violations of scaling diminish only slowly with increasing $\xi$ so giving the semblance that they are negligible. In contrast, for $SU(2) \otimes SU(2)$ finite-size scaling holds with $O(a^2)$ violations and the value deduced for $m/\Lambda_{\overline{MS}}$ agrees well with the prediction.

We also compare the computed mass spectrums in the $SO(N)$ model for $N = 8, 10$ with theory. The interpolating fields for the vector bound states of $SO(N)$ can be easily constructed and $N = 8$ is the smallest value for which there exists more than one such state. The results show a persistent deviation from theory for the range of $\xi$ investigated.

In section 2 we present the relevant 1-loop calculations which relate the various $\Lambda$–parameters; in section 3 we describe the simulation and measurement techniques including a method for variance reduction; in section 4 we give the simulation results and analysis for the study of the $SO(4)$ matrix model; in section 5 we give the simulation results for the $SU(2) \otimes SU(2)$ model; in section 6 we discuss the $N = 6, 8$ models and compare the mass spectrum and $\Lambda_{\overline{MS}}$, computed in the lattice and energy schemes, with theory; in section 7 we discuss our interpretation of all the results and how further studies may elucidate our findings. We also draw our conclusions.

The energy-scheme can be related to the lattice-scheme using the 1-loop calculation for $\langle E \rangle$ expanded in $1/\beta_L \equiv u_L$. This calculation is familiar and we just give the result for general $SO(N)$:

$$\langle E \rangle = \langle 1 - \frac{1}{N} Tr(\mathbf{U}(x,t)\mathbf{U}^T(x+1,t)) \rangle .$$

$$= \frac{N-1}{8\beta_L} \left(1 + \frac{N}{32\beta_L} + \ldots \right) . \quad (2.1.1)$$

The energy-scheme coupling, $\beta_E$, is defined as

$$\beta_E = \frac{N-1}{8\langle E \rangle} , \quad (2.1.2)$$

and the ratio between $\Lambda$-parameters is then

$$\frac{\Lambda_E}{\Lambda_L} = \exp\left(\frac{A}{b_0}\right)$$

$$= \exp\left(\frac{\pi N}{4(N-2)}\right) , \quad (2.1.3)$$

where $\beta_E = \beta_L - A$.

A calculation similar to that described in reference [21] gives the result

$$\frac{\Lambda_{\overline{MS}}}{\Lambda_L} = \sqrt{32} \exp\left(\frac{\pi N}{2(N-2)}\right) . \quad (2.1.4)$$

$\Lambda_\xi$ is related to $\Lambda_L$ by a 1-loop background field calculation in which the 2D theory on a strip is converted to a local 1D theory which can be directly solved for the mass gap since there is an equivalent Schrödinger equation. A similar approach was used in reference [22]. To analyse the 1D system it is sufficient to study the action

$$S(\mathbf{W}) = \beta_{QM} \sum_t Tr(\mathbf{W}_t \mathbf{W}_{t+1}) . \quad (2.1.5)$$

We represent $\mathbf{W}$ by

$$\mathbf{W} = \exp(\boldsymbol{\phi} \cdot \mathbf{T}) ,$$

where the generators $T_i$ are defined so that

$$Tr(T_i T_j) = -\frac{1}{2}\delta_{ij} . \quad (2.1.6)$$

Then for small $\phi$ we have

$$S(\phi) \sim \frac{\beta_{QM}}{4} \sum_t \dot{\phi}^2 + \ldots .$$

This defines the quantum mechanics of a particle constrained to the $SO(N)$ group manifold with mass $\mu = \beta_{QM}/2$. The corresponding Schrödinger equation is then

$$-\frac{1}{2\mu} \mathbf{D}^2 Y_l(\mathbf{W}) = E_l Y_l(\mathbf{W}) . \quad (2.1.7)$$

The $Y_l$ are group harmonics with $l$ a generic label, and satisfy [23]

$$\mathbf{D}^2 Y_l(\mathbf{W}) + C_l Y_l(\mathbf{W}) = 0 \ .$$

Using $Y_0 \propto 1$ and $Y_1(\mathbf{W}) \propto Tr(\mathbf{W})$ we find $C_0 = 0$ and $C_1 = C_F$, the Casimir of the fundamental representation deduced from the generators. From equation (2.1.6) we find

$$\begin{aligned} C_F &= \frac{(N-1)}{4} \\ \Rightarrow & \\ m(L) &= \frac{E_1 - E_0}{2\mu} \\ &= \frac{N-1}{4\beta_{QM}} \ . \end{aligned} \quad (2.1.8)$$

We now calculate $\beta_{QM}$ in terms of $\beta_L$ to 1-loop order. The field variables $\mathbf{U}$ on the 2D strip of width $L$ and length $T$ can be expressed in terms of fluctuations about a background field $\mathbf{W}_t$ which is constant across the strip and slowly varying in $t$, the coordinate along the strip. We write

$$\mathbf{U}(x,t) = \mathbf{W}_t^{1/2} e^{g\phi(x,t)} \mathbf{W}_t^{1/2} \ , \quad (2.1.9)$$

where $\phi(x,t) \equiv \boldsymbol{\phi}(x,t) \cdot \mathbf{T}$ is the fluctuation about $\mathbf{W}_t$. The definition of $W_t$ is given by the relation

$$\sum_x \mathbf{U}(x,t) = \mathbf{W}_t^{1/2} \mathbf{S}_t \mathbf{W}_t^{1/2} \ ,$$

with $\mathbf{S}_t$ a positive symmetric matrix. Using equation (2.1.9) this corresponds at lowest order to the constraint

$$\sum_x \phi(x,t) = 0 \ . \quad (2.1.10)$$

Because $\mathbf{W}_t$ is slowly varying we can expand the effective 1D action in terms of $\Delta_t$ where $\Delta_t \equiv \boldsymbol{\Delta}_t \cdot \mathbf{T}$ and

$$\mathbf{W}_t \mathbf{W}_{t+1}^T = e^{\Delta_t} \ .$$

To 1-loop order this equation can be implemented by choosing

$$\mathbf{W}_t = \mathbf{1} \quad \text{and} \quad \mathbf{W}_{t+1} = e^{-\Delta_t} \ . \quad (2.1.11)$$

Substituting for $\mathbf{U}(x,t)$ and $\mathbf{W}_t$ from equations (2.1.9) and (2.1.11) into the $SO(N)$ action, equation (1.1) we find

$$\begin{aligned} S(\mathbf{U}) &= \beta_L \sum_{x,t} Tr\left(e^{g\phi(x,t)} e^{-g\phi(x+1,t)}\right) \\ &+ \beta_L \sum_{x,t} Tr\left(e^{g\phi(x,t)} e^{-\Delta_t/2} e^{-g\phi(x+1,t)} e^{-\Delta_t/2}\right) \ . \end{aligned} \quad (2.1.12)$$

Expanding to quadratic order, choosing $g^2 = 2/\beta_L$ and using equation (2.1.6) we find

$$S = S_0 + S_I \ ,$$

$$S_I = L\beta_L \sum_t -\tfrac{1}{2}\boldsymbol{\Delta}_t \cdot \boldsymbol{\Delta}_t$$
$$+ \sum_{x,t} Tr\left(2\phi^2\Delta_t^2 + 2\phi'^2\Delta_t^2 - \phi'\phi\Delta_t^2 - \phi\phi'\Delta_t^2 - \phi\Delta_t\phi'\Delta_t\right).$$
$$(2.1.13)$$

The abbreviations $\phi$ for $\phi(x,t)$ and $\phi'$ for $\phi(x,t+1)$ have been used and $\nabla$ is the 2D lattice Laplacian. It should be noted that there is no term **linear** in $\Delta_t$. This is due to the specific form of the decomposition in equation (2.1.9) and the identity

$$Tr\left(\{T_iT_j + T_jT_i\}T_k\right) = 0 \ .$$

The absence of such terms simplifies the calculation since all contributions to the 1D effective action, $S_{\text{eff}}(\Delta)$, are simply given by

$$S_{\text{eff}}(\Delta) = \langle S_I \rangle \ . \tag{2.1.14}$$

Other parametrizations require an evaluation of the quadratic terms in $\langle S_I^2 \rangle$. The average is with respect to the fluctuation measure $\exp(S_0)$ taking into account the constraint in equation (2.1.10) which eliminates the zero mode in the $x$-direction. We use the Gaussian results

$$\langle \phi_i \phi_j \rangle = G(0,0)\,\delta_{ij} \qquad \langle \phi_i \phi'_j \rangle = G(0,1)\,\delta_{ij} \ . \tag{2.1.15}$$

$G(0,0)$ and $G(0,1)$ are given by the expressions

$$G(0,0) = \int_{-\pi}^{\pi} \frac{dq}{2\pi} \sum_{p=1}^{L-1} \frac{1}{4\left(\sin^2(\pi p/L) + \sin^2(q/2)\right)} \ .$$
$$G(0,1) = G(0,0) - \tfrac{1}{4} \ . \tag{2.1.16}$$

We use generators defined in equation (2.1.6) with $k = 1$ and the identities

$$[T_i, T_j] = -f_{ijk}T_k \qquad f_{ijk}f_{ijl} = \frac{N-2}{2}\delta_{ij}$$
$$T_iT_i = \frac{N-1}{4} \qquad Tr(T_iT_jT_k) = -\tfrac{1}{4}f_{ijk} \ . \tag{2.1.17}$$

Then

$$\langle Tr(\phi^2\Delta^2)\rangle = \langle Tr(\phi'^2\Delta^2)\rangle = \frac{N-1}{4}G(0,0)\left(-\tfrac{1}{2}\boldsymbol{\Delta}\cdot\boldsymbol{\Delta}\right) \ ,$$

$$\langle Tr(\phi\phi'\Delta^2)\rangle = \frac{N-1}{4}G(0,1)\left(-\tfrac{1}{2}\boldsymbol{\Delta}\cdot\boldsymbol{\Delta}\right) \ ,$$

$$\langle Tr(\phi\Delta\phi'\Delta)\rangle = \frac{N}{4}G(0,1)\left(-\tfrac{1}{2}\boldsymbol{\Delta}\cdot\boldsymbol{\Delta}\right) \ . \tag{2.1.18}$$

Using equations (2.1.13) to (2.1.18), we find

$$S_{\text{eff}}(\Delta) = L\beta_{\text{eff}} \sum_t \left(-\tfrac{1}{2}\Delta\cdot\Delta\right) , \qquad (2.1.20)$$

where

$$\beta_{\text{eff}} = \beta_L - \frac{N-2}{4}\left(\frac{1}{2\pi}\log L - A\right) - \frac{N}{16} . \qquad (2.1.21)$$

Note that the coefficient of $\log L$ is $b_0$ for the coupling $u_L \equiv 1/\beta_L$. Substituting into equation (2.1.8) with $\beta_{QM} = L\beta_{\text{eff}}$ gives

$$\frac{1}{u(L)} \equiv \frac{1}{m(L)L} = \frac{4\beta_{\text{eff}}}{N-1}$$

$$= \frac{4}{N-1}\left(\beta_L - b_0\left(\log L + \gamma_N\right)\right) ,$$

$$\gamma_N = \frac{\pi N}{2(N-2)} - 2\pi A . \qquad (2.1.22)$$

We thus deduce that

$$\frac{\Lambda_\xi}{\Lambda_L} = \exp(\gamma_N)$$

$$= \exp\left(-2\pi A + \frac{\pi N}{2(N-2)}\right) . \qquad (2.1.23)$$

Using equation (2.1.4) we finally get

$$\frac{\Lambda_{\overline{MS}}}{\Lambda_\xi} = \sqrt{32}\exp(2\pi A) , \qquad (2.1.24)$$

which is independent of $N$.

From equation (2.1.22) we also deduce the tree-level relation

$$u = \frac{N-1}{4}u_L . \qquad (2.1.25)$$

## 2.2 $O(N)$-spin models

The results for the mass-gap for $O(N)$-spin models is given to 3-loop order in [2] and so we include only a brief outline of the 1-loop effective 1D calculation here for completeness. A similar calculation was done by Lüscher [24] but it is instructive to present it in a concise formulation consistent with the previous section. To relate the energy- and lattice-schemes we use the expansion [10]

$$\langle E \rangle = \frac{N-1}{4\beta_L}\left(1 + \frac{1}{8\beta_L}\right) .$$

and the definition

$$\beta_E = \frac{N-1}{4\langle E \rangle}$$

From ref. [21] we have

$$\frac{\Lambda_{\overline{MS}}}{\Lambda_L} = \sqrt{32} \exp\left(\frac{\pi}{2(N-2)}\right) .$$

To calculate $m(L)$ on the $L \times T$ strip we use the same method as in the previous subsection with the $O(N)$ spins $\mathbf{s}(x,t)$ expressed in terms of the background field $\Sigma_t$ and the fluctuation field $\boldsymbol{\phi}(x,t)$ as

$$\mathbf{s}(x,t) = \Sigma_t \sqrt{1 - \boldsymbol{\phi}(x,t) \cdot \boldsymbol{\phi}(x,t)} + \boldsymbol{\phi}(x,t) , \tag{2.2.1}$$

where $\Sigma_t \cdot \Sigma_t = 1$ and $\boldsymbol{\phi}(x,t)$ is an $(N-1)$-dimensional vector satisfying the constraint

$$\sum_x \boldsymbol{\phi}(x,t) = 0 . \tag{2.2.2}$$

This constraint eliminates the zero-mode divergence in the calculation. We then choose

$$\Sigma_t = \Sigma_0 \quad \text{and} \quad \Sigma_{t+1} = e^{\Delta_t T_{12}} \Sigma_0 , \tag{2.2.3}$$

where $\Sigma_0 = (1,0,0,\ldots,0)$ and $\Delta_t$ is the slowly varying 1D background field. $T_{12}$ is the generator of $SO(N)$ rotations in the 12-plane. As in the $SO(N)$ case the particular form of the parametrisation in equation (2.2.1) ensures that the 1D effective action is given by

$$S_{\text{eff}} = \langle S_I \rangle ,$$

there being no linear terms in $\Delta_t$ in the expansion of the $S(\mathbf{s})$ in $\Delta$ and $\boldsymbol{\phi}$. The calculation follows the same steps as in the $SO(N)$ case and gives

$$\beta_{\text{eff}} = \beta_L - (N-2)\left(\frac{1}{2\pi}\log L - A\right) - \tfrac{1}{4} , \tag{2.2.4}$$

where the coefficient of $\log L$ is identified with $b_0$ and, as before, $A = 0.0351637$. Then

$$\frac{\Lambda_\xi}{\Lambda_L} = \exp\left(-2\pi A + \frac{\pi}{2(N-2)}\right) . \tag{2.2.5}$$

Using equation (2.2) we conclude that

$$\frac{\Lambda_{\overline{MS}}}{\Lambda_\xi} = \sqrt{32} \exp(2\pi A) , \tag{2.2.6}$$

which is identical to the $SO(N)$ result. The two results in equations (2.1.24) and (2.2.6) must be the same for $N = 3, 4$ since the two models have the same continuum limit. That they are independent of $N$ ensures the results are identical for all $N$.

The tree-level relation between couplings is then

$$u = \frac{N-1}{2} u_L . \tag{2.2.7}$$

ref. [25] we used an overrelaxation update [26, 27] applied to embedded $O(2)$ models. In terms of CPU-time requirements this algorithm out-performed a multigrid algorithm [28] for correlation lengths up to about 20 in the case of the $CP^4$-model in two dimensions.

In order to save random-numbers, and hence CPU-time, a large fraction of the Metropolis updates have been replaced by a demon-update [29]. Most of the parameters in the algorithm were chosen by trial based on previous experience with the $CP^4$-model [25].

Our basic update steps are performed on $O(2)$ subgroups of the $SO(N)$ group. We have chosen the same subgroup for each of the sites of the lattice. After a number of sweeps a new subgroup is chosen.

The $O(2)$ subgroups that we consider are given by rotations among two rows or two columns.

$$\begin{aligned} U'_{ki}(x,t) &= s_1(x,t)\,U_{ki}(x,t) + s_2(x,t)\,U_{kj}(x,t) \\ U'_{kj}(x,t) &= -s_2(x,t)\,U_{ki}(x,t) + s_1(x,t)\,U_{kj}(x,t) \end{aligned} \qquad (3.1)$$

with $s_1^2 + s_2^2 = 1$.

This parametrization induces an action for the embeded $O(2)$ model

$$S_{cond}(s) = - \sum_{<x,t;x',t'>} \sum_{m,n=1}^{2} c_{mn}(x,t,x',t')\,s_m(x,t)\,s_n(x',t') \qquad (3.2)$$

with

$$\begin{aligned} c_{11} = \phantom{-}c_{22} &= \beta \sum_k U_{ki}(x,t)\,U_{ki}(x',t') + U_{kj}(x,t)\,U_{kj}(x',t') \\ c_{12} = -c_{21} &= \beta \sum_k U_{ki}(x,t)\,U_{kj}(x',t') - U_{kj}(x,t)\,U_{ki}(x',t') \end{aligned} \qquad (3.3)$$

For the updating of the embedded $O(2)$ model we used a combination of standard metropolis, demon-updates and microcanonical updates. We apply the microcanonical update step discussed by [27] for the standard $XY$ model in two dimensions. First we compute the sum of the nearest neighbour spins of the site $(x,t)$.

$$R_m = \sum_{(x',t')nn(x,t)} \sum_{n=1}^{2} c_{mn}(x,t,x',t')\,s_n(x',t') \qquad (3.4)$$

The new values for $\vec{s}$ are then obtained by reflection with respect to $\vec{R}$.

$$\vec{s}\,' = 2\,\frac{\vec{R}\vec{s}}{\vec{R}^2}\,\vec{R} - \vec{s} \qquad (3.5)$$

Since $\vec{R}\vec{s}\,' = \vec{R}\vec{s}$ this update step keeps the action constant.

The aim of the demon-update is to perform updates similar to a Metropolis update but avoiding CPU-intensive parts like the evaluation of trigonometric functions, the exponential function and pseudo-random numbers.

The demons are introduced by an additional term in the action

$$S' = S + \sum d_{x,t} \qquad (3.6)$$

applied to the demons.

$$d = -\ln(\eta), \tag{3.7}$$

where $\eta$ is a pseudo-random number with a uniform distribution in the interval $]0,1]$. Then we perform updates that keep the composite action of the spin model plus the demons constant, exchanging energy between the demons and the spins. First we compute a proposal for a new spin-value $\vec{s}\,'$ by reflecting $\vec{s}$ at the sum of the upper and left neighbour spins. Then we check whether the demon at the site can take over the energy without becoming negative. If this is the case, we accept the proposal $\vec{s}\,'$ and set the demon to it's new value. After a sweep of such demon updates we translate the demons on the lattice.

For one given embedding we performed 1 standard Metropolis sweep, and 10 to 60 overrelaxation sweeps, and finally 5 demon-updates. The number of overrelaxation sweeps was chosen to be roughly proportional to the correlation length.

We alternate the row embedding and the column embedding. The first raw (column) is chosen in a fixed sequence from 1 to N, while the second was chosen randomly from the remaining ones.

## 3.1 The improved correlation function estimator

In order to obtain a meaningful result for the step-scaling function the correlation length on the finite lattices has to be computed with an accuracy of less then one percent. In order to achieve this aim we used the improved estimator for the correlation function discussed in ref. [30] for the case of $O(N)$ vector models.

The underlying physical idea for this improved estimator is similar as that for the 1-loop solution of the model discussed in section 2. However here instead of an **effective** one dimensional model we rather use a **conditional** (or embedded) model. For a given field configuration $U$ the conditional model is defined by

$$S_{cond}(W) = -\beta \sum_t \sum_x Tr\left(W(t)\ U(x,t)\ (W(t+1)\ U(x,t+1))^T\right) \tag{3.1.1}$$

where $W$ is the field of the conditional model. Note that there is no spatial part in the action, since $W(t)$ does not depend on $x$. Performing the $x$ summation we get

$$S_{cond}(W) = -\sum_t Tr\left(Q(t,t+1)\ W^T(t+1)\ W(t)\right) \tag{3.1.2}$$

where

$$Q(t,t+1) = -\beta \sum_x U(x,t)\ U^T(x,t+1)\ . \tag{3.1.3}$$

Reparametrising the model by $R(t,t+1) = W^T(t)W(t+1)$ we obtain

$$S_{cond}(R) = -\sum_t Tr\left(Q(t,t+1)\ R^T(t,t+1)\right)\ . \tag{3.1.4}$$

For free boundary conditions in time-direction there is no constraint on the $R^T(t, t+1)$. Therefore the partition function factorizes, and the solution of the conditional 1-D model is reduced to the solution of zero-dimensional systems. The conditional expectation value of the time-slice correlation function is given by

$$\langle G(t,t+\tau)\rangle_{cond} = Tr\ (S(t)\ \langle R(t,t+1)\rangle_{cond}...\langle R(t+\tau-1,t+\tau)\rangle_{cond}\ S(t+\tau)) \tag{3.1.5}$$

We were not able to compute the conditional expectation values exactly. Instead we used Monte Carlo integration for this task. Firstly one has to note that finite statistics for the conditional expectation value does not corrupt the end-result for $\langle G(t, t+\tau)\rangle$. Secondly a enormous gain in statistical accuracy can be obtained, since as a consequence of the factorization in eq. (3.1.5) the statistics of the single "baby" Monte Carlo's multiply. Typically we performed 200 Metropolis update steps for the evaluation of $\langle R(t, t+1)\rangle_{cond}$. We used the final value of $R$ for updating the field $U$. However, we have yet to make careful tests to determine the efficiency gain of this measurement technique.

## 4 Simulation results for the $SO(4)$ model

The main object of this investigation was to test the predictions for $m/\Lambda_{\overline{MS}}$ for the $SO(N)$ principal chiral models given by Hollowood [1, 20]. The action for the $SO(N)$ matrix models is given in equation (1.1). It was found that even for moderate values of $N$ ($N \geq 6$) the continuum limit was difficult to achieve with any degree of confidence, all indications being that a large correlation length is necessary. This is in contrast to recent work on $SU(N)$ models [8, 9] where the results indicate that the models are close to the continuum limit even for small $\xi$ ($\xi > 5$). In the next section we will present some results for $N = 6$, 8 and 10 to support these statements but will postpone our speculations concerning why such a difficulty occurs until the conclusion.

In this section we concentrate on the $SO(4)$ model and apply the renormalization scheme described in [2], the '$\xi$-scheme', and compare it with the lattice- and energy-schemes used in [8, 9]. The coupling constant $u(Lm_\infty)$ is defined in equation (1.3) [24] and to achieve the continuum limit we require $\bar{u}(Lm_\infty, a/L)$ for fixed $Lm_\infty$ to be essentially independent of $a$ and hence that $\xi_\infty$ is large enough that the $a$-dependent corrections to equation (1.3) are negligible. The problem is that in practice this might require $\xi_\infty$ to be very large indeed and unachievable in a present-day simulation. Alternatively, it might be possible to fit the $a$-dependence in equation (1.3) by measuring $\bar{u}(Lm_\infty, a/L)$ for various $\xi_\infty$ with $Lm_\infty$ fixed, and extrapolating to $a/L = 0$. In the case that the corrections are perturbative they behave as $O(a^2)$. Whether or not this is the case must be deduced from the simulation and for the $SO(4)$ model there is clear evidence that a simple perturbative interpretation of the $a$-dependent effects is not possible for the values of $\xi_\infty$ we use. Nevertheless we try to extrapolate the results to the continuum limit in a reliable way and compute the value of $\bar{u}(2, a/L)$. It must be emphasized that it is crucial to determine this value of $u$ as accurately as possible since it is the starting point for the subsequent determination of $u$ at smaller scales, and ultimately contributes to the systematic error on the computation of $m/\Lambda_{\overline{MS}}$. In order to attain the continuum limit we measured $u(2)$ for increasing values of $\xi_\infty$ and fitted the $a$-dependence and so deduced the continuum coupling, $u(2)$. We characterised the theory either by the lattice coupling $\beta_L$ or by the coupling defined by Parisi [19] in terms of the internal energy, equation (2.1.1):

$$\beta_E = \frac{N-1}{8\langle E\rangle} \ . \tag{4.1}$$

Note $\langle E\rangle$ is normalized to lie in $[0, 1]$. We found that $\langle E\rangle$ suffered from strong finite-size effects and so could not be expressed as a function of $\beta_L$ alone. We chose to measure by trial we can see which method gives the better extrapolation to $a/L = 0$. Where $L/a = 2\xi_\infty$ was non-integral, simulations were done for the integer $L/a$ values either side of $2\xi_\infty$ and interpolation used to deduce $\xi(L)$. The results are shown in table 1 where we can see that the energy-method shows better convergence than the lattice-method. The errors shown are statistical and a naive straightforward extrapolation gives the value $u(2) = 2.25(2)$ for the continuum coupling constant. However, we shall argue below that whilst plausible this value for $u(2)$ is incorrect since the convergence to the continuum is only apparent not real.

| $\beta_L$ | $\langle E \rangle$ | $\xi_\infty$ | $\bar{u}(2, a/L)$ lattice-method | $\bar{u}(2, a/L)$ energy-method |
|---|---|---|---|---|
| 1.05 | 0.5411(2) | 3.71(3) | 1.98(1) | 2.44(1) |
| 1.10 | 0.4702(2) | 8.34(7) | 2.07(1) | 2.35(2) |
| 1.12 | 0.4435(1) | 13.62(13) | 2.11(1) | 2.27(2) |
| 1.14 | 0.42176(6) | 25.3(4) | 2.17(2) | 2.26(2) |

Table 1: The values of $\bar{u}(2, a/L)$ in lattice- and energy-methods as a function of $\xi_\infty$. The energy-method shows plausible convergence to the extrapolated value $u(2) = 2.25(2)$.

That $\langle E \rangle$ is sensitive to the lattice size is another indication that $SO(N)$ models are more complex in the continuum limit than $SU(N)$ models where no such effect is observed. We find that the finite-size effects in $\langle E \rangle$ largely seem to offset those in $u(Lm_\infty)$ which indicates that there is some connection between the short and long-range properties of the system. From table 1 it can also be seen that the $a$-dependent effects in the lattice-method are not fitted by a perturbative parametrization: they are closer to a $a^{1/2}$ dependence. This indicates that non-perturbative contributions are strong and that their effect on the calculation of $u(2)$ seems to be largely accounted for using the energy-method. These are all reasons why we should be cautious and suspicious of assuming that we are observing properties of the continuum theory.

One reasonable test that we are simulating the continuum theory is to check that the computed value of $m/\Lambda_{\overline{MS}}$ agrees with the theoretical prediction. We thus compute the step-scaling function described in [2] and section 1 and fit the short-distance behaviour of $u$ to the form deduced from perturbation theory. We choose the factor for the scale-change to be $s = 1/2$. Therefore we consider pairs of lattices with sizes $2L$ and $L$ respectively and $\beta_L$ is adjusted so that $\tilde{u}(2L/a, \beta_L)$ is a required value. Then $\tilde{u}(L/a, \beta_L)$ is measured in both the lattice and energy schemes, that is, keeping either $\beta_L$ or $\beta_E$ constant on the lattice pair. Various values of $L/a$ were chosen so that the $a$-dependent effects can be determined and eliminated by extrapolation. When we believe that the continuum limit has been attained then the step scaling function $\sigma$ can be determined from equation (1.5) and compared with perturbation theory using equations (1.6,1.7)

$$b_0 = \frac{(N-2)}{2\pi(N-1)} \qquad b_1 = \frac{b_0^2}{2} .$$

The first two coefficients of the $\beta$-function quoted for $u$ are obtained from those asso-

| | | | | lattice-method | energy-method |
|---|---|---|---|---|---|
| 8  | 1.0908  | 0.4531(5)   | 2.25   | 1.590(10) | 1.727(10) |
| 14 | 1.1169  | 0.4352(3)   | 2.25   | 1.647(8)  | 1.720(8)  |
| 26 | 1.1393  | 0.4186(2)   | 2.25   | 1.711(7)  | 1.737(7)  |
| 40 | 1.15335 | 0.40915(7)  | 2.25   | 1.734(7)  | 1.747(7)  |
| 7  | 1.1088  | 0.4347(3)   | 1.747  | 1.416(3)  | 1.475(3)  |
| 13 | 1.1347  | 0.4194(2)   | 1.747  | 1.469(3)  | 1.480(3)  |
| 20 | 1.15235 | 0.4082(1)   | 1.747  | 1.492(3)  | 1.499(3)  |
| 30 | 1.16917 | 0.39778(6)  | 1.747  | 1.506(3)  | 1.509(3)  |
| 10 | 1.1484  | 0.4111(1)   | 1.517  | 1.323(2)  | – |
| 15 | 1.1659  | 0.3989(1)   | 1.517  | 1.340(3)  | – |
| 20 | 1.1803  | 0.3911(1)   | 1.517  | 1.345(2)  | – |
| 5  | 1.1424  | 0.4074(3)   | 1.351  | 1.203(2)  | – |
| 10 | 1.1789  | 0.3906(2)   | 1.351  | 1.212(2)  | – |
| 16 | 1.2045  | 0.3784(1)   | 1.351  | 1.218(2)  | – |
| 8  | 1.2028  | 0.3774(2)   | 1.222  | 1.113(1)  | – |
| 16 | 1.245   | 0.36014(9)  | 1.222  | 1.114(1)  | – |
| 8  | 1.2452  | 0.3587(2)   | 1.115  | 1.025(1)  | – |
| 16 | 1.2914  | 0.34215(8)  | 1.115  | 1.024(1)  | – |
| 8  | 1.2922  | 0.3408(2)   | 1.0234 | 0.9494(7) | – |
| 16 | 1.3368  | 0.32681(7)  | 1.0234 | 0.954(1)  | – |
| 8  | 1.3357  | 0.3264(1)   | 0.948  | 0.8902(5) | – |
| 16 | 1.3851  | 0.31230(6)  | 0.948  | 0.8910(4) | – |
| 8  | 1.3849  | 0.3116(1)   | 0.891  | 0.8355(5) | – |
| 16 | 1.4356  | 0.29861(6)  | 0.891  | 0.8342(5) | – |

Table 2: Couplings measured in the lattice- and energy-methods for different values of $L/a$. The $a$-dependent violations to finite-size scaling are most pronounced for $Lm_\infty \sim 1$. The energy-method was not used where it gave little information in addition to the lattice-method. The statistical errors of $\tilde{u}(2L/a)$ which are not quoted here are of similar size as those of $\tilde{u}(L/a)$. coupling values.

ciated with $u_L = 1/\beta_L$ [21] by using equation (2.1.25) gives the tree-level relation

$$u = \frac{(N-1)}{4} u_L .$$

The results are shown in table 2.

In each case the continuum coupling for a given scale was determined by extrapolation to $a = 0$. From table 2 it can be seen that the most care needs to be taken when $Lm_\infty \sim 1$ where the corrections to finite-size scaling are the greatest. In all cases the energy-method was the most convergent, but for sufficiently large $L$ both the lattice- and energy-methods were compatible and a common value for the continuum coupling was consistent. For $Lm_\infty < 1/2$ the $a$−dependent corrections were small and consistent with perturbation theory and both schemes gave consistent results. Following eq. (1.8) we tried to use the result of a given step as the argument of the next step. This was achieved in all cases except when $Lm_\infty = 1/2$ and $Lm_\infty = 1/128$ where twe corrected the small mismatches by interpolation. These corrections lead to scale changes of 2.026 and 1.840 respectively. The sequence of continuum couplings deduced are shown in table 3.

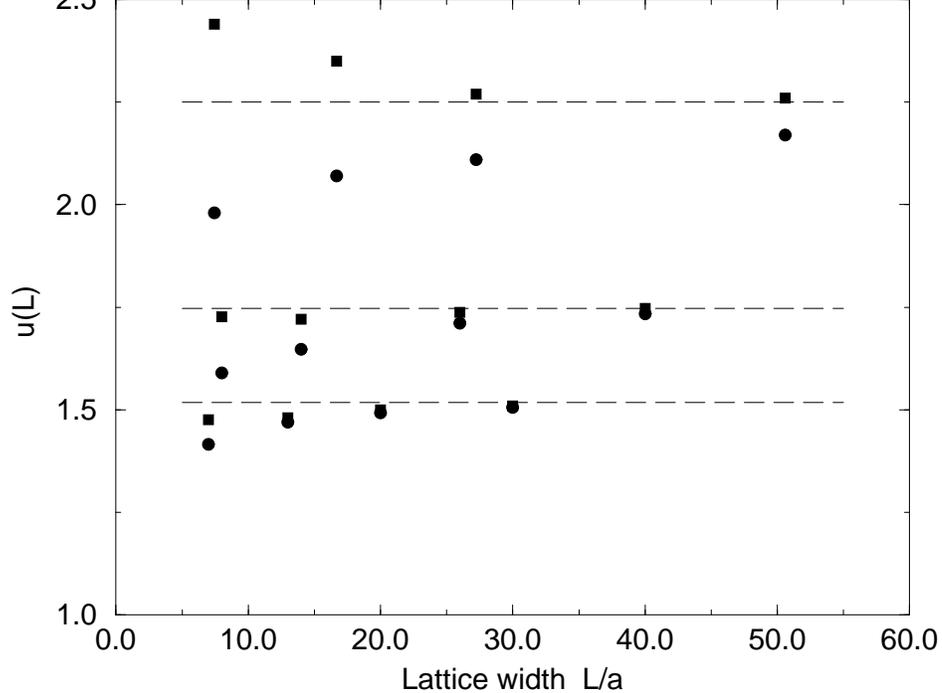

Figure 1: Values of the coupling $\bar{u}(Lm_\infty, a/L)$ from table 2 plotted against $L/a$ $Lm_\infty$ fixed at the values $2, 1, 1/2$. The violations of the finite-size scaling assumption are clearly evident for both versions of the $\xi$-scheme used: the lattice-scheme (●) and the energy-scheme (■). In each case the value assumed for the extrapolation to $a = 0$ is shown as the dashed line

Also shown are the 1− and 2−loop approximations to $\sigma(1/2, u)$ evaluated with $u = u(2Lm_\infty)$. These should be compared with the Monte-Carlo result

$$\sigma(1/2, u(2Lm_\infty))_{MC} = u(Lm_\infty) .$$

Clearly asymptotic scaling sets in for $Lm_\infty \leq 1/4$. It is perhaps surprising that the 2−loop approximation fits the Monte-Carlo results so well at such relatively large scales, but it gives confidence that we can probe deeply into the region where asymptotic scaling holds and hence that perturbative parametrization of $u$ in terms of $\Lambda_{\overline{MS}}$ is valid. The $a$-dependence of the results for these smaller values of $u$ indicates that $\beta_L$ is large enough for the extrapolation to the continuum limit, $a = 0$, to be reliable. The data from table 3 are shown in figure 2 where the computed function $\sigma(2, u)$ is compared with its 1− and 2−loop approximations deduced from equations (1.6,1.7). The important intermediate region where it is crucial to maintain the continuum limit in order to reliably relate the low and high energy scales is also in clear evidence. This match between the small-$L$ region and large-$L$ region (where trivially $u(Lm_\infty) = Lm_\infty$) is the important part of the simulation.

From table 3 we can deduce the value of $m/\Lambda_\xi$, i.e., in the $\xi$-scheme, using the 2−loop formula, equation (1.9). We have:

$$\Lambda_\xi^{(2)} = \frac{1}{L} \left(\frac{u}{3\pi}\right)^{-\frac{1}{2}} \exp\left(-\frac{3\pi}{u}\right) . \tag{4.2}$$

|           |         |             | 1-loop  | 2-loop | 3-loop |
|-----------|---------|-------------|---------|--------|--------|
| 1         | 2.25    | 1.754(13)   | 1.93054 | 1.9010 | 1.8915 |
| .987/2    | 1.747   | 1.517(6)    | 1.5481  | 1.5329 | 1.5291 |
| .987/4    | 1.517   | 1.351(6)    | 1.3647  | 1.3544 | 1.3521 |
| .987/8    | 1.351   | 1.222(4)    | 1.2289  | 1.2214 | 1.2199 |
| .987/16   | 1.222   | 1.1148(21)  | 1.1212  | 1.1154 | 1.1145 |
| .987/32   | 1.1148  | 1.0234(15)  | 1.0303  | 1.0259 | 1.0252 |
| .987/64   | 1.0234  | 0.9478(11)  | 0.9518  | 0.9483 | 0.9478 |
| 1.073/128 | 0.9561  | 0.8912(12)  | 0.8933  | 0.8904 | 0.8900 |
| 1.073/256 | 0.8912  | 0.8339(12)  | 0.8364  | 0.8340 | 0.8337 |

Table 3: The sequence of continuum couplings as a function of $Lm_\infty$. Also shown are the 1− and 2−loop approximations to $\sigma(1/2,u)$ evaluated with $u = u(2Lm_\infty)$. These should be compared with the Monte-Carlo result $\sigma(1/2,u(2Lm_\infty))_{MC} = u(Lm_\infty)$. It seems plausible that asymptotic scaling sets in for $Lm_\infty \leq 1/4$.

In fact, since the covering group of $SO(4)$ is $SU(2) \otimes SU(2)$ and the $SU(2)$ matrix model is isomorphic to the $O(4)$ spin model, we are able to use the 3-loop $\beta$-function in reference [2] to deduce that

$$\Lambda_\xi^{(3)} = \Lambda_\xi^{(2)} \left(1 - \frac{1}{6\pi}u\right) \quad . \tag{4.3}$$

A discussion of the detailed relationship between these models will be postponed until the next section but it is convenient to invoke this formula here. The 1-loop calculation necessary to determine the ratio $\Lambda_\xi/\Lambda_{\overline{MS}}$ is given in section 2, and using the result for $\Lambda_L/\Lambda_\xi$ and $\Lambda_{\overline{MS}}/\Lambda_L$ from equations (2.1.4) and (2.1.23), we give the computed 2- and 3-loop results for $m/\Lambda_{\overline{MS}}$ versus length scale in table 4. The prediction for this quantity can be taken from the paper by Hollowood [1] where the formula for $m/\Lambda_{\overline{MS}}$ for the $SO(N)$ matrix models can be extended down to $N = 4$ [20]. The prediction is

$$\frac{m}{\Lambda_{\overline{MS}}} = \frac{2^{7/2}}{\sqrt{\pi e}} = 3.8715 \quad . \tag{4.4}$$

| $Lm_\infty$ | 1/256 | 1/128 | 1/64 | 1/32 | 1/16 | 1/8 | 1/4 | 1/2 | 1/1 |
|---|---|---|---|---|---|---|---|---|---|
| $m/\Lambda_{\overline{MS}}$ 2-loop | 14.3(1.0) | 14.3(1.0) | 14.4(9) | 14.4(9) | 14.1(9) | 14.1(9) | 14.2(9) | 14.0(7) | 13.2(5) |
| $m/\Lambda_{\overline{MS}}$ 3-loop | 13.7(1.0) | 13.6(1.0) | 13.7(9) | 13.6(9) | 13.3(9) | 13.2(9) | 13.2(9) | 12.9(7) | 12.0(5) |

Table 4: Computed values for $m/\Lambda_{\overline{MS}}$ using the 2-loop $\beta-$function for the coupling evaluated at the indicated scale (the small adjustments to the scale explicit in table 3 are omitted for clarity). The errors are calculated from the accumulation of statistical errors at all preceeding steps.

As can be seen from table 4 the computer results are wrong by a minimum of a factor of four. The inescapable conclusion is that we have not eliminated corrections to finite-size scaling and are not close to the continuum limit in our simulation. Without the theoretical prediction we might have been persuaded that the evidence did point to the simulation having produced reliable continuum results − the step scaling function strongly violated is for lattices where $Lm_\infty \sim 1$ and the parts of the simulation which are important for setting the scale are not probing the continuum. The corollary is that we need the inital value of $\xi_\infty = 1/am_\infty$ to be considerably larger than the biggest we have taken. The surmise is that since the $SO(N)$ manifold is not simply connected, $\Pi_1(SO(N)) = Z_2$, there are vortices which are responsible for non-perturbative violations of finite-size scaling and much larger values of $\beta$ are needed before their effect is sufficiently suppressed so that the continuum limit can be approached in a controlled, perturbative, way. One check is therefore to simulate the covering group of $SO(4)$, $SU(2) \otimes SU(2)$, and see whether the problems with finite-size scaling violations are perturbative. The results are presented in the next section.

We can also check what happens if we apply the method used in references [12, 10, 8, 9] for the couplings defined in equation (1.11). These bare couplings are defined to be evaluated at the scale of the lattice spacing $a$. The $\Lambda$-parameters of the respective lattice- and energy-schemes, $\Lambda_L$ and $\Lambda_E$ can be calculated using equation (1.9). Since the corresponding value of $am_\infty$ is known from the simulation the estimate for $m/\Lambda_{\overline{MS}}$ can be calculated using equations (2.1.3) and (2.1.4). The results are shown in table 5. The results are very poor and it is clear that the approach is hopeless. This should be contrasted with the success of the method applied to $SU(N)$ matrix models [8, 9] and the $O(4)$ and $O(8)$ spin models [10].

| $\beta_L$ | $\langle E \rangle$ | $\xi_\infty$ | $m/\Lambda_{\overline{MS}}$ | |
|---|---|---|---|---|
| | | | lattice-scheme | energy-scheme |
| 1.05 | 0.5411(2) | 3.71(3) | 305(2) | 20.4(2) |
| 1.10 | 0.4702(2) | 8.34(7) | 248(2) | 31.3(3) |
| 1.12 | 0.4435(1) | 13.62(13) | 194(2) | 34.1(3) |
| 1.14 | 0.42176(6) | 25.3(4) | 133(2) | 30.9(5) |
| 1.15 | 0.41291(5) | 33.4(8) | 114(3) | 29.5(7) |

Table 5: The values of $m/\Lambda_{\overline{MS}}$ in the lattice- and energy-schemes [12, 10, 8, 9] as a function of $\xi_\infty$ for the $SO(4)$ model. The theoretical prediction is $m/\Lambda_{\overline{MS}} = 3.87153$ [1, 20]

## 5  Results for the $SU(2) \otimes SU(2)$ covering of $SO(4)$

The $SU(2) \otimes SU(2)$ matrix model consists of two independent $SU(2)$ models and the $SU(2)$ matrix model is isomorphic to the $O(4)$ spin model where the fields take values in $O(4)/O(3) \equiv S^3$. Thus we need only to simulate the $O(4)$ spin model which can be done using the cluster algorithm [11]. The theoretical prediction for $m/\Lambda_{\overline{MS}}$ is given in [7, 6] to be

$$m/\Lambda_{\overline{MS}} = \sqrt{\frac{32}{\pi e}} = 1.9358 \ . \qquad (5.1)$$

The only difference between the $SO(4)$ and $SU(2)$ matrix models is that the lightest state of the $SU(2)$ model is a spinor and that of the $SO(4)$ model corresponds to a spinor-antispinor state which is not, in fact, bound. Thus, in the $SO(4)$ model the large time asymptotics of the correlator are controlled by the spinor-antispinor cut and not a bound state pole. However, in two dimensions the large time behaviour is dominated

$$m_{SO(4)} = 2m_{SU(2)} . \tag{5.2}$$

This factor of two simply converts the $SO(4)$ prediction of Hollowood [1] into the prediction of Hasenfratz et al. [7, 6] for the $O(4)$ spin model. It then follows that the two continuum couplings defined as in equation (1.3) are related by a factor of two:

$$u_{SO(4)} = 2u_{SU(2)} . \tag{5.3}$$

We shall omit the distinguishing subscript on $u$ unless it is necessary to avoid ambiguity.

The action for the $O(4)$ spin model is taken to be

$$S(\mathbf{s}) = \beta_L \sum_{n,\mu} \mathbf{s}_n \cdot \mathbf{s}_{n+\mu} . \tag{5.4}$$

As before we define $u$ using equations (1.2) and (1.3). The tree-level result from section 2.2 relating the $\xi$- and lattice-schemes is

$$u = \frac{N-1}{2} u_L . \tag{5.5}$$

Using this result, the $\beta$−function for $u$ as defined in equation (1.7) then has coefficients

$$b_0 = \frac{(N-2)}{\pi(N-1)} , \quad b_1 = \frac{N-2}{\pi^2(N-2)^2} , \quad b_2 = \frac{N-2}{\pi^3(N-2)^2} . \tag{5.6}$$

The result for $b_2$ is taken from reference [2]. In addition to the 2-loop formula for $\Lambda_\xi$, equation (1.9), we have the 3-loop formula

$$\Lambda_\xi^{(3)} = \Lambda_\xi^{(2)} \left( 1 - \frac{1}{\pi(N-1)} u \right) . \tag{5.7}$$

Because the continuum limits of the $SO(4)$ and $O(4)$ spin model are controlled by the same Lie algebra the conversion ratio $\Lambda_\xi/\Lambda_{\overline{MS}}$ is the same for both. However, as a check and for completeness, the 1-loop calculation which yields this conversion ratio for the general $O(N)$ spin model is briefly described in section 2.2. Of course, for $N = 4$ all necessary results can be taken from reference ([2]).

| $\beta_L$ | $\langle E \rangle$ | $\chi$ | $\xi_\infty$ | $\bar{u}(1, 1/\xi_\infty)$ |
|---|---|---|---|---|
| 2.00 | 0.4230(2) | 99.8(3) | 7.89(2) | 1.552(5) |
| 2.20 | 0.3775(1) | 266.6(8) | 13.95(5) | 1.562(4) |
| 2.40 | 0..34070(5) | 555.7(1.9) | 25.7(1) | 1.585(5) |
| 2.60 | 0.31071(2) | 1629.9(2.6) | 47.07(8) | 1.584(3) |

Table 6: The values of $\bar{u}(1, 1/\xi_\infty)$ in the lattice-method as a function of $\xi_\infty$. Interpolation is used to calculate $\bar{u}$ at non-integer values of $L$. Clearly, finite-size scaling violations are small and, within the errors, not incompatible with the perturbative prediction that they behave like $O(a^2)$

The simulation results for $u(1)$ for various values of $\xi_\infty$ are shown in tables 6 and 7 and the results for the corresponding step-scaling function are given in table 8 where the values quoted have been extrapolated to the $a = 0$ limit. Clearly the violations of finite-size scaling are much smaller than for the $SO(N)$ matrix model, compare with tables 1 and 2, and are compatible, within errors, with the perturbative prediction that they behave like $O(a^2)$. In comparing tables 7 and 2 the conversion factor of two between couplings, eqn (5.3) should be bourne in mind. Note that we have not used the energy-method for the spin model since it is not needed.

| | | | | | |
|---|---|---|---|---|---|
| 7 | 2.1914 | 0.3714(2) | 47.87(6) | 1.584(4) | 1.210(3) |
| 13 | 2.4032 | 0.3378(1) | 140.8(2) | 1.584(3) | 1.219(2) |
| 24 | 2.6050 | 0.30941(3) | 415.3(4) | 1.584(3) | 1.227(1) |
| 8 | 2.4649 | 0.3264(1) | 78.0(1) | 1.288(3) | 1.008(2) |
| 16 | 2.6959 | 0.29735(5) | 268.0(3) | 1.288(2) | 1.011(2) |
| 8 | 2.6954 | 0.29579(7) | 96.5(1) | 1.011(2) | 0.865(1) |
| 16 | 2.9260 | 0.27185(4) | 337.9(4) | 1.011(2) | 0.863(1) |

Table 7: Couplings measured in the lattice-method for different values of $L/a$. The $a$-dependent violations of finite-size scaling are apparent but small, even for the crucial case where $L/a \sim \xi_\infty$

| $Lm_\infty$ | $u(2Lm_\infty)$ | $u(Lm_\infty)$ | $\sigma(1/2, u(2Lm_\infty))$ | | |
|---|---|---|---|---|---|
| | | | 1-loop | 2-loop | 3-loop |
| 2/1 | 4.132(10) | 2.309(10) | 2.5700 | 2.2927 | 2.1062 |
| 1/1 | 2.309(10) | 1.584(4) | 1.7236 | 1.6384 | 1.5925 |
| 1/2 | 1.584(4) | 1.228(3) | 1.2847 | 1.2495 | 1.2348 |
| 1/4 | 1.228(2) | 1.011(3) | 1.0401 | 1.0216 | 1.0152 |
| 1/8 | 1.011(2) | 0.863(2) | 0.8801 | 0.8689 | 0.8657 |

Table 8: The sequence of continuum couplings for the $O(4)$ spin model as a function of $Lm_\infty$. Also shown are the 1− and 2−loop approximations to $\sigma(1/2, u)$ evaluated with $u = u(2Lm_\infty)$. These should be compared with the Monte-Carlo result $\sigma(1/2, u(2Lm_\infty))_{MC} = u(Lm_\infty)$. Clearly, asymptotic scaling is already setting in for $Lm_\infty \leq 1$

The 2-loop and 3-loop computed values of $m/\Lambda_{\overline{MS}}$ are given in table 9. The 2-loop result is already near to the predicted value of 1.9358 for $Lm_\infty \sim 1/8$ and the 3-loop result agrees with this prediction within errors even for $Lm_\infty \sim 1$. In figure 2 we compare the $SU(2) \otimes SU(2)$ and $SO(4)$ results for the step-scaling function. We have plotted $u_{SO(4)}$ and $2u_{SU(2)}$ on the ordinate since in the continuum limit the data points should coincide up to the small correction factors of table 3. Also shown are the curves for $\sigma(1/2, u)$ derived from the 1-loop and 3-loop $\beta$-functions. The large $u$ result, $\sigma(1/2, u) \to u/2$, to which all curves should eventually be asymptotic is also shown. It can be seen that there is a clear deviation between the results for the two models in the

| $L/a\xi_\infty$ | 1/8 | 1/4 | 1/2 | 1/1 |
|---|---|---|---|---|
| $m/\Lambda_{\overline{MS}}$ 2-loop | 1.783(30) | 1.735(28) | 1.679(18) | 1.609(10) |
| $m/\Lambda_{\overline{MS}}$ 3-loop | 1.963(33) | 1.944(31) | 1.930(20) | 1.935(11) |

Table 9: Computed values for $m/\Lambda_{\overline{MS}}$ using the 2-loop and 3-loop $\beta$−functions for the coupling evaluated at the indicated scale. The errors are calculated from the accumulation of statistical errors at all preceeding steps. These results are to be compared with the theoretical prediction [7, 6] $m/\Lambda_{\overline{MS}} = 1.9358$

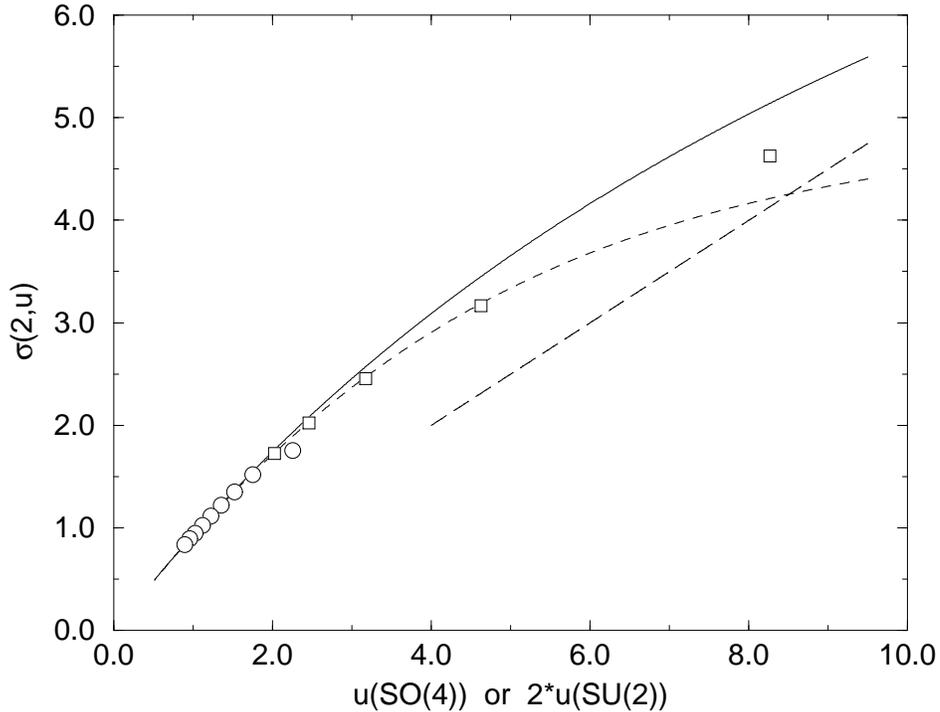

Figure 2: The step-scaling function $\sigma(1/2, u)$ versus $u$. The solid and dashed curves are respectively the 1- and 3-loop perturbative calculations. Data for the $SO(4)$ matrix model (○) and the $SU(2) \otimes SU(2)$ matrix model (□). The large $u$ result to which all curves are asymptotic is shown as the long-dashed line.

The calculations of $m/\Lambda_{\overline{MS}}$ in the lattice- and energy-schemes have been done using 3-loop results by Wolff in ref. [10]. Our simulation results agree in detail with his and he finds general agreement with the prediction for $m/\Lambda_{\overline{MS}}$. However, the different schemes tried by Wolff do show different rates of convergence to theory as a function of $\xi_\infty$ the best being the energy-scheme which agrees very well with theory for $\xi_\infty \sim 10$. We have similarly good agreement for the $\xi$-scheme confirming the ease with which the continuum limit can be controlled.

## 6  Results for $SO(N)$, $N = 6, 8, 10$

We attempted to analyse the $SO(N)$ models for $N = 6, 8, 10$ in the same way as for $SO(4)$. However, the CPU time required is prohibitively large and we were unable to work with sufficiently large correlation lengths. It is instructive, however, to compute the values of $m/\Lambda_{\overline{MS}}$ for each model in the lattice- and energy-schemes. The results are shown in table 10. Whilst the energy-scheme does not show the large deviation from theory of the $SO(4)$ model the results are clearly untrustworthy. There is no agreement between the different schemes and, although the range of $\xi$ used was limited, there is for $N = 8, 10$ the energy-scheme gives results which seem independent of $\xi_\infty$ for the restricted range covered. Clearly, this cannot be taken as indicating that the results have converged to the $\xi_\infty \to \infty$ limit: in light of our experience it shows very little.

Another prediction derived from the exact $S$-matrix is for the mass-ratios of particles in the theory. For $SO(N)$ ($N$ even) the prediction is

$$\frac{m_p}{m} = \frac{\sin(\pi p/(N-2))}{\sin(\pi/(N-2))} \qquad 1 < p \leq (N-2)/2 \;, \tag{6.1}$$

where $p$ labels the $p$-th species and $m$ is the mass of the lightest state. We simulated the $SO(N)$ model for $N = 6, 8, 10$ and measured the masses in the different channels labelled by $p$. Hollowood [1] has discussed the relevant interpolating operators for these states and we choose the simplest operators which couple to the desired state in each channel. The operator for the $p$-th state is

$$O_p(\mathbf{x})_{i_1\ldots i_p j_1\ldots j_p} = \sum_{perms\ of\ j_1\ldots j_p} (-1)^P U(\mathbf{x})_{i_1 j_1}\ldots U(\mathbf{x})_{i_p j_p} \;, \tag{6.2}$$

where $P$ is the permutation signature of the ordering of the $\{j_i\}$. Thus $O_p$ is the outer product of $p$ matrices antisymmetrized on the row and column labels respectively. The corresponding Green function is

$$G_p(\mathbf{x}) = \langle Tr(O_p(\mathbf{x}) O_p^T(0)) \rangle_c \;, \tag{6.3}$$

where the trace has the obvious meaning. The results for the mass ratios $m_2/m$ for $N = 8, 10$ and $m_3/m$ for $N = 10$ are shown in table 11.

| $N$ | $\beta_L$ | $\langle E \rangle$ | $\xi_\infty$ | $m/\Lambda_{\overline{MS}}$ | | |
|---|---|---|---|---|---|---|
| | | | | lattice-scheme | energy-scheme | theory |
| 6 | 1.5 | 0.6786(1) | 1.387(3) | 48.8(1) | 5.32(1) | 3.87153 |
| 6 | 1.6 | 0.6226(1) | 1.97(1) | 62.3(3) | 6.03(3) | 3.87153 |
| 6 | 1.7 | 0.5181(3) | 4.85(4) | 46.0(4) | 7.98(7) | 3.87153 |
| 6 | 1.73 | 0.4784(2) | 8.7(1) | 30.7(3) | 8.07(9) | 3.87153 |
| 8 | 2.28 | 0.5727(3) | 2.80(2) | 35.3(3) | 5.27(4) | 3.65837 |
| 8 | 2.31 | 0.5362(4) | 3.98(3) | 28.0(2) | 5.55(4) | 3.65837 |
| 8 | 2.33 | 0.5073(3) | 5.75(7) | 21.0(3) | 5.51(7) | 3.65837 |
| 8 | 2.35 | 0.4851(2) | 8.05(6) | 16.2(1) | 5.36(4) | 3.65837 |
| 10 | 2.91 | 0.5623(2) | 3.05(1) | 25.13(8) | 4.65(2) | 3.523789 |
| 10 | 2.93 | 0.5341(4) | 4.18(3) | 19.46(14) | 4.61(3) | 3.523789 |
| 10 | 2.95 | 0.5054(2) | 5.90(3) | 14.63(7) | 4.62(2) | 3.523789 |

Table 10: The values of $m/\Lambda_{\overline{MS}}$ computed in the lattice- and energy-schemes compared with the theoretical prediction for $N = 6, 8, 10$. There is no agreement between the schemes and no trend suggesting that the results will converge to the prediction.

There is no convincing agreement between simulation and theory and, moreover, no trend suggesting that the discrepancy is $\sim a^2$.

## 7 Discussion

The main result of this paper is that the properties of the continuum $SO(4)$ theory cannot be observed in a simulation of the lattice-regularized model for the values of $\beta_L$ and correlation lengths accessible to current computers. We have shown that no such difficulty occurs for the model based on the $SU(2) \otimes SU(2)$ cover of $SO(4)$. We believe that both models give rise to the same continuum theory, characterized by the fixed point at $\beta_L = \infty$, but that the ways in which this continuum theory is approached in the lattice-regularized versions are very different. For $SU(2) \otimes SU(2)$ finite-size scaling with $O(a^2)$ deviations holds for the range of couplings used, and for scales $Lm_\infty < 1$ the flow with $L$ of the renormalized coupling, $u(Lm_\infty)$, is well given by the 3-loop $\beta$-function. The value for $m/\Lambda_{\overline{MS}}$ computed in the simulation agrees well with the theoretical prediction (table 9). In contrast, for $SO(4)$ the violations of scaling do not fit an $O(a^2)$ form but do seem to diminish to zero (see fig. 1) as $\xi_\infty$ increases. This apparent or 'pseudo-scaling' can be wrongly interpreted as signalling the continuum theory and can deceive us into believing that the value for the continuum coupling can be deduced. The false nature of this pseudo-scaling is exposed by comparing the resulting computed value for $m/\Lambda_{\overline{MS}}$ of $\sim 14$ with the theoretical prediction of 3.8715. We also find no convergence to a result for $m/\Lambda_{\overline{MS}}$ using the lattice- and energy- schemes for $SO(4)$. In contrast for $SU(2) \otimes SU(2)$ the lattice-scheme gives an acceptable result although clearly inferior to that of the $\xi$-scheme.

| | | | simulation | theory | simulation | theory |
|---|---|---|---|---|---|---|
| 8 | 2.28 | 2.80(2) | 1.84(1) | 1.732 | - | - |
| 8 | 2.31 | 3.98(3) | 1.79(3) | 1.732 | - | - |
| 8 | 2.33 | 5.75(7) | 1.83(2) | 1.732 | - | - |
| 8 | 2.35 | 8.05(6) | 1.85(2) | 1.732 | - | - |
| 10 | 2.91 | 3.05(1) | 1.89(2) | 1.848 | 2.7(2) | 2.414 |
| 10 | 2.93 | 4.18(3) | 1.93(2) | 1.848 | 2.73(3) | 2.414 |
| 10 | 2.95 | 5.90(3) | 1.91(2) | 1.848 | 2.73(3) | 2.414 |

Table 11: The computed mass-ratios for $N = 8, 10$ compared with the predictions. There is no convincing agreement between simulation and theory.

We analysed data for the $SO(N)$ models with $N = 6, 8, 10$ and used the lattice- and energy-schemes to attempt to obtain an estimate for $m\Lambda_{\overline{MS}}$. The results are shown in table 10 and it is clear that there is no agreement between the schemes nor with the theoretical prediction. We also computed the mass-ratios of the fundamental masses predicted by Ogievetsky et al. [5] and the results are given in table 11. There is a persistent discrepancy up to 10% and there is no sign of a trend to the correct values as $\beta_L$ increases. These results support the conjecture that we are unable to simulate the continuum theory for $SO(N)$ models with present computer resources.

The discrepancy between the computed and theoretical values of $m/\Lambda_{\overline{MS}}$ is much larger than that of about 10% reported by Lüscher et al. in their analysis of the $O(3)$ spin model [2], which they attributed to the truncation of the perturbative $\beta$-function at 3-loop order. This is not the resolution of the problem we have found for the $SO(4)$ model. We conjecture that the difference lies in the different connectivities of the underlying manifolds: the $SO(N)$ models contain $Z_2$ vortex lattice artifacts whilst the covering group models do not. In the general $SO(N)$ case the cover is $Spin(N)$ which is constructed from the associated Clifford algebra [31]. The vortices create an obstruction to observing the fixed point at $\beta_L = \infty$ but are eventually suppressed at sufficiently large $\beta_L$. Recent work by Hasenbusch [16] and Niedermayer et al. [17] has discussed a similar phenomenon comparing the $O(3)$ and $RP^2$ spin models. They propose a similar conclusion, namely that the difference between the two models is due to the nls interpretation that true scaling has set in. From fig. 1 we might confidently deduce that $u(2) = 2.25$ but from our simulation of the covering group we find that the value should be $u = 3.17$. We expect that as $\beta_L$ increases a cross-over phenomenon will occur where the violations to scaling will again become large and then eventually diminish to become $O(a^2)$ allowing the true scaling limit to appear. It is important to estimate the value of $\beta_L$ and $\xi_\infty$ above which finite-size scaling and the continuum theory limit should be observed. A crude attempt can be made with current data by using the theoretical prediction for $m/\Lambda_{\overline{MS}}$ together with

$$\xi_\infty = \frac{L/a}{m/\Lambda_\xi}(b_0 u(Lm_\infty))^{1/2} \exp(\frac{1}{b_0 u(Lm_\infty)}) \; , \qquad (7.1)$$

to deduce $\xi_\infty(\beta_L)$ from the data for $u(Lm_\infty)$ at sufficiently small $Lm_\infty$ given in table 3. These results for $\xi_\infty(\beta_L)$ can be compared with the results for $\xi_\infty(\beta_L)$ computed directly from simulation. If the lattice theory is near the continuum limit these alternative methods of computing $\xi_\infty(\beta_L)$ should give similar answers. There is consistency to within 10% for the values $\xi_\infty(\beta_L)$ from using similar values of $\beta_L$ on lattices of different widths corresponding to different measured values of $\tilde{u}$. Where there is a choice we have taken the result from the smallest $\tilde{u}$. We plot $\log(\xi_\infty)$ versus $\beta_L$ in figure 3 for both approaches. The mismatch is clear for $\beta_L \sim 1.15$: the direct measurement gives $\xi_\infty = 33.4(8)$ whereas the short-distance result is $\xi_\infty \sim 150$. It seems reasonable to infer that the two methods will not agree until $\xi_\infty \gg 150$.

The perturbative step-scaling function is well reproduced for sufficiently small scales even though the observables on scales $Lm_\infty \sim 1$ show a large departure from continuum behaviour. In this context note that $\sigma(1/2, u)$ for $u = 2.25$ is actually quite close to the 3-loop prediction even though the true scale associated with this value is a factor of about four different from that assigned in the simulation. Thus near agreement with the perturbative prediction at small scales is not sufficient to infer that the continuum theory is being observed at large scales: from our simulation of the covering group the correct value for $u$ at the scale assigned in this case is $u = 3.17$ and $\sigma(1/2, u)$ for this value agrees very well indeed with the 3-loop perturbative prediction (see fig 2).

Suppose we were able to simulate at, say, $\beta_L \sim 1.18$ on a large enough lattice. From our results we see that the properties of the continuum theory are well reproduced at scales $Lm_\infty < 1/8$ but from the above discussion we would expect the direct measurement of $\xi_\infty$ to be considerably less than the short-distance prediction of $\xi_\infty \sim 300$. So whilst the properties of the continuum theory are computable at short distances for $\beta_L \sim 1.18$ the long distance results do not reflect the continuum but are dominated by residual lattice artifacts: vortices in the $SO(4)$ model.

For an euclidian continuum field theory a field configuration can be viewed as a map of the space $R^2$ onto the manifold of the field. A vortex is now characterized by the property that the map of a loop in $R^2$ onto the manifold of the field is not smoothly contractible. As a consequence there is at least one sigularity of the field inside such a loop. In statisitical mechanics, vortices have been mainly discussed in relation with the two dimensional $XY$ model. The classical energy of a vortex is given by $E \approx \pi \; \log(R/a)$ where $R$ is the size of the vortex. Based on the simple energy versus entrophy argument that the free energy is given by $F = E - TS$ with $S = 2 \log(R/a)$, Kosterlitz and Thouless [33] inferred the occurrence of a phase transition at $T = \pi/2$.

This argument does rely on the assumptions, that the free energy of a vortex at a fixed location is essentially given by its energy and that knowledge of the free-energy a fixed location is bounded as $R \to \infty$ for non-abelian theories [32] and so a simple KT style analysis cannot be carried out. However, it has also been suggested [34, 32] that for non-abelian theories the interaction between vortices is such that the free-energy of multi-vortex configurations cannot be inferred from the properties of an isolated vortex. It is clear that further studies are necessary to clarify the true position.

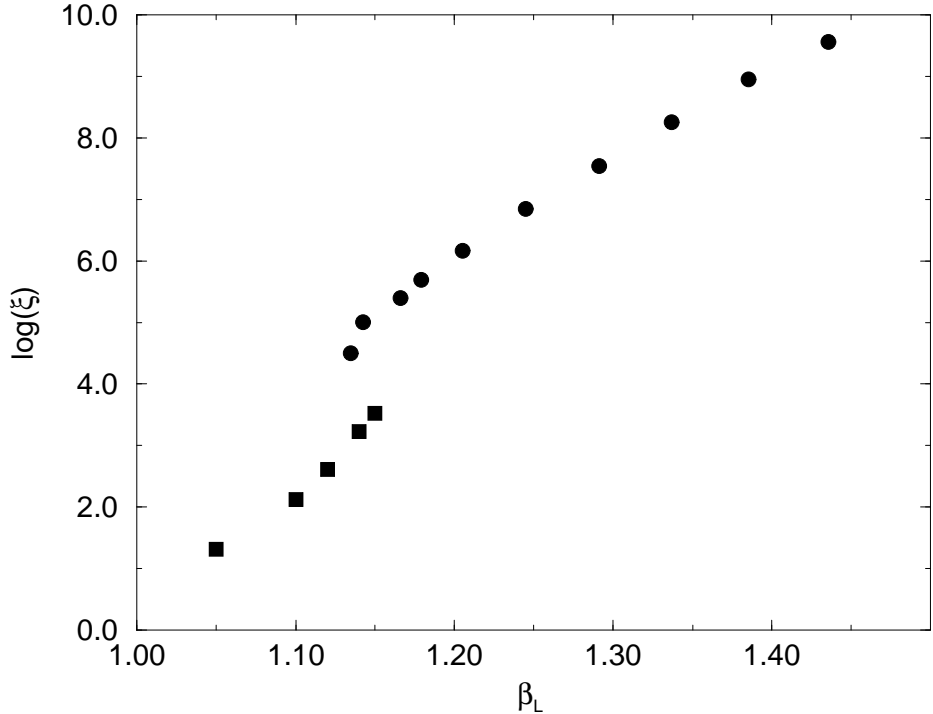

Figure 3: $\log(\xi_\infty)$ plotted versus $\beta_L$ derived from the short-distance behaviour and the 2-loop $\beta$-function (●) and from direct measurement in the simulation (■). The two sets of results do not agree indicating that the large scale properties are not controlled by the continuum theory.

A number of approaches can be taken:

(i) The free-energy of an isolated vortex can be calculated at least in 1-loop perturbation theory and in a lattice simulation to check that the argument above can be made. Also the free-energy of multi-vortex configurations should be computed by simulation;

(ii) The vortices of $SO(N)$ can be suppressed by eliminating any configuration which contains one or more vortices. The finite-size scaling analysis can be repeated to see whether the continuum theory is more readily observed;

(iii) The matrix models based on $Spin(N)$ can be simulated and compared with the $SU(N)$ models and the mass-ratios predicted from the exact $S$-matrix can be computed;

$$S(s, \mathbf{s}) = \beta \sum_{x,t} \sigma_1(x,t)\mathbf{s}(x,t) \cdot \mathbf{s}(x+1,t) + \sigma_2(x,t)\mathbf{s}(x,t) \cdot \mathbf{s}(x,t+1)$$
$$- \log(z) \sum_{plaquettes} P(\sigma) , \qquad (7.2)$$

where $\mathbf{s}$ is an $N$−component vector of unit length, $\sigma_\mu$, $\mu = 1, 2$, is a gauge field taking values in $[1, -1]$ and $P(\sigma)$ signifies the plaquette.

(b) The model whose manifold has the same topology as the $SO(4)$ manifold but in which the vortex operators can be explicitly constructed. This model has action

$$S(\sigma, \mathbf{s}, \mathbf{r}) = \beta \sum_{x,t} \sigma_1(x,t) \left( \mathbf{s}(x,t) \cdot \mathbf{s}(x+1,t) + \mathbf{r}(x,t) \cdot \mathbf{r}(x+1,t) \right)$$
$$+ \sigma_2(x,t) \left( \mathbf{s}(x,t) \cdot \mathbf{s}(x,t+1) + \mathbf{s}(x,t) \cdot \mathbf{s}(x,t+1) \right)$$
$$- \log(z) \sum_{plaquettes} P(\sigma) , \qquad (7.3)$$

where $\mathbf{s}$ and $\mathbf{r}$ are four-component vectors of unit length and $\sigma_\mu$ and $P(\sigma)$ are as defined in (a).

In both models the renormalization group flow in $(\beta, z)$ can be studied using Monte-Carlo methods and hence the effect of vortices, measured by $P(\sigma)$ and controlled by the fugacity, $z$, can be determined.

These projects are currently in hand.

The outcome is that we should be wary of claims that the continuum theory has been observed which are based on the observation of scaling in a limited window in the coupling constant. Even if properties of the continuum are observed at short distances it does not follow that observables on the scale of the correlation length, which are sensitive to so-called non-perturbative effects, are controlled by the continuum theory. This could be the case for any lattice model which has non-trivial topological artifacts. It has been pointed out [35] that QCD is such a theory since the gauge group is $SU(3)/Z_3$ where $Z_3$ is the centre of $SU(3)$ [36]. It is important to determine whether such an effect exists in QCD and at what level of accuracy it needs to be taken into account in present-day simulations. We have been unable to observe any continuum properties in the $SO(N)$ matrix models but it remains to be seen whether $Spin(N)$ models have the same problems.

# 8   Acknowledgements

We wish to thank Dr. I.T. Drummond, Dr. F. Niedermayer and Professor U. Wolff for useful discussions and advice. We also wish to thank the Leverhulme Trust for the award of a grant to support this work and MH wishes to thank them for the award of a Leverhulme Fellowship. RRH thanks the Royal Society for the award of a research grant for equipment on which much of the computer simulation was done.

[2] M. Lüscher, P. Weisz, and U. Wolff. *Nucl. Phys.*, B359:221–243, 1991.

[3] A.B. Zamolodchikov and Al.B. Zamolodchikov. *Ann. of Phys.*, 120:25, 1979.

[4] A.M. Polyakov and P.B. Wiegmann. *Phys. Lett.*, 131B:121, 1983.

[5] E. Ogievetsky et al. *Nucl. Phys.*, B280[FS 18]:45–96, 1987.

[6] P. Hasenfratz and F. Niedermeyer. *Phys. Lett.*, 245B:529–532, 1990.

[7] P. Hasenfratz, M. Maggiore, and F. Niedermeyer. *Phys. Lett.*, 245B:522–528, 1990.

[8] P. Rossi and E Vicari. *Phys. Rev.*, D49:1621–1628, 1994.

[9] I.T. Drummond and R.R. Horgan. *Phys. Lett.*, 321B:246–253, 1994.

[10] U. Wolff. *Phys. Lett.*, 248B:335–339, 1990.

[11] U. Wolff. *Nucl. Phys.*, B334:581–610, 1994.

[12] U. Wolff. *Phys. Lett.*, 222B:473–475, 1989.

[13] S. Caracciolo et al. *Nucl. Phys. B (Proc. Suppl.)*, 30:815, 1993.

[14] S. Caracciolo et al. *Phys. Rev. Lett.*, 71:3906, 1993.

[15] S. Caracciolo et al. *Nucl. Phys. B (Proc. Suppl.*, 34:129, 1994.

[16] M. Hasenbusch. 1995. University of Cambridge, DAMTP preprint 95/??

[17] N. Niedermayer, P. Weisz, and Dong-Shin Shin. 1995. Bern/Max Planck preprint BUTP-95-19/MPI-PHT-95-55.

[18] M.P. Nightingale. *J. Appl. Phys.*, 53:7927, 1982.

[19] G. Parisi. in Proc.XXth Conf. on High Energy Physics (Madison,WI, 1980).

[20] T.J. Hollowood. Private communication.

[21] J. Shigemitsu and J.B. Kogut. *Nucl. Phys.*, B190[FS3]:365–411, 1981.

[22] P. Hasenfratz and F. Niedermayer. *Z. Phys.*, B92:91, 1993.

[23] I.T. Drummond, S Duane, and R.R. Horgan. *Nucl. Phys.*, B220:119, 1983.

[24] M. Lüscher. *Phys. Lett.*, 118B:391, 1982.

[25] M. Hasenbusch and S. Meyer. *Phys. Rev. D*, 45:4376–4380, 1992.

[26] S.L. Adler. *Phys. Rev. D*, 23:2901, 1981.

[27] R. Gupta et al. *Phys. Rev. Lett.*, 68:1996, 1988.

[28] M. Hasenbusch and S. Meyer. *Phys. Rev. Lett.*, 68:435–438, 1992.

[29] M. Creutz. *Phys. Rev. Lett.*, 50:1411, 1983.

[32] A. Sokal. *Private Communication.*

[33] J.M. Kosterlitz and D.J. Thouless. *J. Phys.*, C6:1181, 1973.

[34] F. Niedermayer. *Private Communication.*

[35] S. Solomon, Y. Stavans, and E. Domany. *Phys. Lett.*, 112B:373–378, 1982.

[36] G. t'Hooft. *Nucl. Phys.*, B138:1, 1978.